\documentclass[journal]{IEEEtran}
\usepackage{amsmath,amsfonts}
\usepackage{algorithm2e}
\usepackage{array}
\usepackage[caption=false,font=normalsize,labelfont=sf,textfont=sf]{subfig}
\usepackage{textcomp}
\usepackage{stfloats}
\usepackage{url}
\usepackage{verbatim}
\usepackage{graphicx}
\usepackage{cite}

\usepackage{makecell}

\usepackage{graphicx} 
\usepackage{amsmath}
\usepackage{amsmath}
\usepackage{amsmath, amssymb}
\usepackage{xcolor}	

\usepackage{booktabs}

\usepackage{algpseudocode}
\usepackage{graphics}
\usepackage{amsmath, amsthm}

\usepackage[caption=false,font=normalsize,labelfont=sf,textfont=sf]{subfig}

\newcounter{assumption}

\newcounter{theorm} 

\begin{document}
	
	\title{Pinching Antenna System–Assisted Hybrid AirComp–NOMA Uplink: Joint Precoding and Antenna Placement Optimization
    }


	\author{IEEE Publication Technology,~\IEEEmembership{Staff,~IEEE,}
	}

	\author{Saeid Pakravan, Imene Trigui, Wessam Ajib, Wei-Ping Zhu 

\thanks{S. Pakravan and W. Ajib are with the Department of Computer Sciences, University of Quebec in Montreal (UQAM), Montreal, QC, Canada. email: pakravan.saeid@uqam.ca; ajib.wessam@uqam.ca.}

\thanks{I. Trigui is with the Department of Applied Sciences, University of Quebec at Chicoutimi (UQAC), Chicoutimi, QC, Canada. email: itrigui@uqac.ca.}

\thanks{W.-P. Zhu is with the Department of Electrical and Computer Engineering, Concordia University, Montreal, QC, Canada. email: weiping@ece.concordia.ca.}


}
	\maketitle

\begin{abstract}

This paper studies a pinching antenna system (PAS)-assisted hybrid uplink architecture that integrates over-the-air computation (AirComp) and non-orthogonal multiple access (NOMA) to simultaneously support distributed data aggregation and individual communication services. A base station with a dielectric waveguide hosting multiple pinching antennas receives signals from AirComp and NOMA users over shared time–frequency resources. To assess joint computation–communication performance, a hybrid metric combining the AirComp computation rate and the NOMA sum rate is proposed. Based on this metric, a joint optimization problem is formulated to maximize the hybrid rate by optimizing user transmit precoding, receive combining, and antenna deployment, subject to power, quality-of-service, and aggregation accuracy constraints. An alternating optimization framework is developed to solve the resulting non-convex problem. Numerical results show that the proposed design achieves significant performance gains over several benchmark schemes.


\end{abstract}


\section{Introduction}

The rapid evolution toward sixth-generation (6G) wireless networks, coupled with the explosive growth of the Internet of Things (IoT), is driving unprecedented demands for ultra-low latency, massive connectivity, and high spectral efficiency \cite{10054381}. Emerging 6G applications, such as immersive extended reality, autonomous systems, and large-scale distributed intelligence, demand not only reliable data delivery but also native support for computation within the wireless infrastructure \cite{9144301}. Conventional transmit-then-compute paradigms, where data are first decoded and then processed, face fundamental scalability and latency limitations in dense multiple-access scenarios.

Over-the-air computation (AirComp) has recently emerged as a promising solution. By exploiting the natural superposition property of the wireless multiple-access channel (MAC), AirComp enables the computation of functions of distributed data directly in the air \cite{Wang2024}. This simultaneous transmission and aggregation approach reduces communication latency and signaling overhead, making it attractive for federated learning, distributed sensing, and edge intelligence \cite{11111711, 9535447, pakravan2024robust}. In parallel, non-orthogonal multiple access (NOMA) has been widely studied to enhance spectral efficiency and support massive connectivity, allowing multiple users to share the same time–frequency resources via power-domain multiplexing and successive interference cancellation (SIC) \cite{8603758}.

Despite their advantages, AirComp and NOMA have largely been studied in isolation, as they are traditionally designed to serve different objectives. Specifically, AirComp focuses on the accuracy of function computation, whereas NOMA aims to maximize individual data rates. In future 6G networks, however, heterogeneous service coexistence is unavoidable, where computation-oriented devices (e.g., sensing and learning nodes) and communication-oriented users (e.g., mobile broadband terminals) operate simultaneously over shared wireless resources \cite{8869705}. This coexistence introduces challenges in interference management, resource allocation, and the tradeoff between computation fidelity and communication throughput.

Recent advances in reconfigurable antenna technologies offer additional spatial degrees of freedom (DoF) to address these challenges. Among these technologies, pinching antenna systems (PAS) have emerged as a novel architecture in which the effective antenna position can be continuously adjusted along a conductive or dielectric medium via controlled pinching mechanisms \cite{10945421,
11263923, 11123791}. Unlike fixed-position antennas, PAS can dynamically adapt to channel conditions and user distributions. This flexibility enables joint spatial and signal-domain optimization, which is especially beneficial in interference-limited integrated communication–computation systems.

This paper studied a PAS-enabled uplink framework supporting two types of users: AirComp users, which transmit function-related signals for over-the-air aggregation, and NOMA users, which seek reliable and high-rate data transmission. A base station (BS) equipped with PAS simultaneously aggregates AirComp signals while decoding individual data streams from NOMA users over shared spectrum resources. The spatial flexibility offered by PAS is leveraged to reshape the effective channels, thereby balancing the conflicting requirements of computation accuracy and communication rate.

The main objective of this work is to optimize overall system performance using a hybrid rate metric that jointly captures NOMA data rates and AirComp computation efficiency. The resulting optimization problem is highly non-convex, involving coupled variables such as antenna pinching positions, transmit power allocation, and receiver design, under practical constraints like power budgets and decoding requirements. To solve this, we develop an alternating optimization (AO) framework that decomposes the original problem into tractable subproblems, solved iteratively. Numerical results show that the proposed design significantly outperforms benchmark schemes, achieving higher hybrid rates while providing valuable insights into the tradeoff between computation accuracy and communication throughput.

\section{System Model and Problem Formulation}

\subsection{System Model}

We consider a PAS-aided uplink multiple-access system, where the BS is equipped with a dielectric waveguide hosting $N$ PAs. The system simultaneously serves two disjoint classes of single-antenna users: a set of AirComp users denoted by $\mathcal{K}_{\mathrm{A}} \triangleq \{1,\ldots,K_{\mathrm{A}}\}$ and a set of uplink NOMA users denoted by $\mathcal{K}_{\mathrm{N}} \triangleq \{1,\ldots,K_{\mathrm{N}}\}$, with $\mathcal{K}_{\mathrm{A}} \cap \mathcal{K}_{\mathrm{N}} = \emptyset$. The overall user set is given by $\mathcal{U} \triangleq \mathcal{K}_{\mathrm{A}} \cup \mathcal{K}_{\mathrm{N}}$. The BS aggregates wireless data from the AirComp users while decoding individual data streams from NOMA users over the same time–frequency resources.
We adopt a three-dimensional (3D) Cartesian coordinate system. The users are uniformly distributed over a rectangular area on the $x$--$y$ plane with side lengths $L_x$ and $L_y$. The position of user $u \in \mathcal{U}$ is denoted by $\boldsymbol{u}_u = (x_u, y_u, 0)$.
Without loss of generality, the dielectric waveguide is deployed along the $x$-axis at a height $d$ above the ground, with length $L_x$. The position of the $n$-th PA on the waveguide is given by $\boldsymbol{v}_n = (v_n, 0, d), \ 0 \leq v_n \leq L_x$, while the feed point of the waveguide is located at
$\boldsymbol{v}_0 = (0, 0, d)$.

The channel coefficient between user $u$ and the $n$-th PA is modeled as \cite{11206485}
\begin{equation}
h_{u,n}
= \frac{\lambda}{4\pi \|\boldsymbol{u}_u - \boldsymbol{v}_n\|}
\exp\!\left(
-j\frac{2\pi}{\lambda}\|\boldsymbol{u}_u - \boldsymbol{v}_n\|
\right),
\end{equation}
where $\lambda$ denotes the carrier wavelength.

Let $s_k \in \mathbb{C}$ and $d_j \in \mathbb{C}$ represent the data symbols transmitted by AirComp user $k \in \mathcal{K}_{\mathrm{A}}$ and NOMA user $j \in \mathcal{K}_{\mathrm{N}}$, respectively. Accordingly, the received signal at the $n$-th PA is expressed as
\begin{equation}
r_n =
\sum_{k \in \mathcal{K}_{\mathrm{A}}}
h_{k,n} p_k s_k
+
\sum_{j \in \mathcal{K}_{\mathrm{N}}}
h_{j,n} p_j d_j
+ z_n,
\end{equation}
where $p_k$ and $p_j$ are the transmit precoding coefficients of AirComp and NOMA users, respectively, and $z_n \sim \mathcal{CN}(0,\sigma^2)$ denotes the additive white Gaussian noise (AWGN).

For analytical tractability, all symbols are assumed to be zero-mean with unit power, i.e.,
$\mathbb{E}[s_k] = 0,  \mathbb{E}[|s_k|^2] = 1,  \forall k \in \mathcal{K}_{\mathrm{A}}, \
\mathbb{E}[d_j] = 0,  \mathbb{E}[|d_j|^2] = 1,  \forall j \in \mathcal{K}_{\mathrm{N}}$.
Furthermore, the transmitted symbols are mutually independent,
$\mathbb{E}[s_k s_i^{{H}}] = 0,  \forall k \neq i, \
\mathbb{E}[d_j d_i^{{H}}] = 0,  \forall j \neq i$. Then, the per-user transmit power constraint is expressed as
\begin{equation}
\mathbb{E}\!\left[ \left| p_u x_u \right|^2 \right]
= |p_u|^2 \leq P_u, \quad \forall u \in \mathcal{U},
\end{equation}
where $x_u \in \mathbb{C}$ denotes the information symbol of user $u$, i.e., $x_u = s_k$ for $u \in \mathcal{K}_{\mathrm{A}}$ and $x_u = d_j$ for $u \in \mathcal{K}_{\mathrm{N}}$, and $P_u > 0$ represents the maximum transmit power budget of user $u$.

Assuming a fully passive waveguide structure, the aggregated signal observed at the feed point is given by
\begin{equation}
\tilde{r}
=
\sum_{n=1}^{N}
r_n
\exp\!\left(
-j\frac{2\pi n_{\mathrm{ref}}}{\lambda}
\|\boldsymbol{v}_n - \boldsymbol{v}_0\|
\right),
\end{equation}
where $n_{\mathrm{ref}}$ denotes the refractive index of the waveguide. Defining the equivalent uplink channel as
\begin{equation}
g_u \triangleq
\sum_{n=1}^{N}
h_{u,n}
\exp\!\left(
-j\frac{2\pi n_{\mathrm{ref}}}{\lambda}
\|\boldsymbol{v}_n - \boldsymbol{v}_0\|
\right),
\end{equation}
the received signal can be compactly written as
\begin{equation}
\tilde{r}
=
\sum_{k \in \mathcal{K}_{\mathrm{A}}}
g_k p_k s_k
+
\sum_{j \in \mathcal{K}_{\mathrm{N}}}
g_j p_j d_j
+ \tilde{z},
\end{equation}
where $\tilde{z} \sim \mathcal{CN}(0, N\sigma^2)$.

\subsection{NOMA Decoding via SIC and Achievable Rates}

Since all users are served over the same bandwidth simultaneously, interference management is a critical issue in the hybrid network composed of both AirComp and NOMA users, particularly regarding inter-type interference. To mitigate this, the BS employs SIC at the feed point of the PAS. It first decodes the signals of strong users in a successive order to remove their interference, and subsequently obtains the superposed signal of weaker users. 
 Specifically, the NOMA users are indexed such that
    $|g_1|^2 \le |g_2|^2 \le \cdots \le |g_{K_{\mathrm{N}}}|^2$, where user~$1$ has the weakest channel and is decoded first, and user~$K_{\mathrm{N}}$ has the strongest channel and is decoded last. At the decoding of user~$j$, signals from users $\{1,\ldots,j-1\}$ are assumed to be perfectly canceled via SIC, whereas signals from users $\{j+1,\ldots,K_{\mathrm{N}}\}$ remain as residual intra-NOMA interference. 
Moreover, AirComp users transmit functional data symbols intended for over-the-air aggregation rather than individual messages. Since AirComp signals cannot be canceled via SIC and thus remain as non-removable interference at all NOMA decoding stages, the BS adopts the proposed decoding order to suppress intra-NOMA interference while accounting for the unavoidable AirComp interference.

Based on the above SIC structure, the signal-to-interference-plus-noise ratio (SINR) for decoding the signal of NOMA user~$j$ is given by
\begin{equation}
\Gamma_j
=
\frac{|g_j|^2 |p_j|^2}
{\sum\limits_{t=j+1}^{K_{\mathrm{N}}} |g_t|^2 |p_t|^2
+
\sum\limits_{k\in\mathcal{K}_{\mathrm{A}}} |g_k|^2 |p_k|^2
+
N\sigma^2 }.
\label{eq:noma_sinr_nopi}
\end{equation}

Accordingly, the achievable uplink rate of NOMA user~$j$ is 
\begin{equation}
R_{{N},j}
=B
\log_2\!\left(1+\Gamma_j\right),
\label{eq:noma_rate_nopi}
\end{equation}
where $B$ denotes the system bandwidth and the NOMA sum-rate is given by
\begin{equation}
R_{N}
=
\sum_{j=1}^{K_{\mathrm{N}}} R_{N,j}.
\end{equation}

\subsection{AirComp Estimation and Computation Rate}

When all signals from NOMA users have been decoded via SIC technique, the BS subtracts them from the received superposed signal to obtain the aggregated signal of all AirComp users for model aggregation, which can be expressed as
\begin{equation}
\hat{r}
=
\sum_{k \in \mathcal{K}_{\mathrm{A}}}
g_k p_k s_k
+ \tilde{z}.
\end{equation}
The goal of AirComp is to estimate $s \triangleq \sum_{k \in \mathcal{K}_{\mathrm{A}}} s_k$.
The BS applies a scalar receive combining coefficient $w\in\mathbb{C}$ to obtain
\begin{equation}
\hat{s}
=
w^{H}
\sum_{k \in \mathcal{K}_{\mathrm{A}}}
g_k p_k s_k
+
w^{H}\tilde{z}.
\end{equation}
The computation distortion of $\hat{s}$ with respect to $s$ is quantified by the MSE, defined as $\mathrm{MSE}(\hat{s}, s) \triangleq \mathbb{E}\left[ \left| \hat{s} - s \right|^2 \right]$,
given by
\begin{equation}
    {\mathrm{MSE}(\hat{s},s)}=
\sum_{k \in \mathcal{K}_{\mathrm{A}}}
\left| w^{H} g_k p_k - 1 \right|^2
+
N\sigma^2 | w |^2.
\label{eq:aircomp_mse}
\end{equation}
 Since it directly measures the computation distortion, it serves as a key performance metric for collaborative computation. Accordingly, the achievable computation rate is defined as \cite{94147855}
\begin{equation}
R_{{A}}
\triangleq
B \log_2\!\left(
1 +
\frac{\mathbb{E}\!\left[|\hat{s}|^2\right]-\mathrm{MSE}(\hat{s},s)}
{\mathrm{MSE}(\hat{s},s)}
\right).
\label{eq:aircomp_rate_def}
\end{equation}

\subsection{Hybrid Performance Metric}

To jointly capture AirComp accuracy and NOMA throughput, we define the hybrid rate as
\begin{equation}
R_{{H}}
\triangleq
\alpha R_{{A}}
+
(1-\alpha)R_{N},
\label{eq:hybrid_rate_final}
\end{equation}
where $\alpha \in [0,1]$ is a weighting parameter that controls the relative importance of AirComp aggregation fidelity and NOMA communication throughput. 
The proposed hybrid metric enables unified evaluation of heterogeneous uplink services and reduces to AirComp optimization or conventional NOMA sum-rate maximization when $\alpha=1$ or $\alpha=0$, respectively.

\subsection{Problem Formulation}

We aim to maximize the achievable hybrid rate in \eqref{eq:hybrid_rate_final} by jointly optimizing the users' transmit precoding coefficients, the deployment of the PAs along the dielectric waveguide, and the receive combining coefficient at the BS. Accordingly, the hybrid performance maximization problem is formulated as
\begin{subequations}
\label{eq:opt_problem_final}
\begin{align}
\max_{w,\,\boldsymbol{p}_k,\,\boldsymbol{p}_j,\,\boldsymbol{v}} \quad
& R_{{H}}(w,\boldsymbol{p}_k,\boldsymbol{p}_j,\boldsymbol{v}) \label{eq:opt_obj} \\
\mathrm{s.t.}\quad
& |p_k|^2 \le P_k, \quad \forall k \in \mathcal{K}_{\mathrm{A}}, \label{eq:aircomp_power_constraint} \\
& |p_j|^2 \le P_j, \quad \forall j \in \mathcal{K}_{\mathrm{N}}, \label{eq:noma_power_constraint} \\
& R_{N,j} \ge R_{\min,j}, \quad \forall j \in \mathcal{K}_{\mathrm{N}}, \label{eq:noma_rate_constraint} \\
& \mathrm{MSE}(\hat{s},s) \le \varepsilon_0, \label{eq:aircomp_mse_constraint} \\
& 0 \le v_n \le L_x, \quad \forall n=1,\ldots,N, \label{eq:pa_constraint} 
\end{align}
\end{subequations}
where $\boldsymbol{p}_k \triangleq [p_k]_{k\in\mathcal{K}_{\mathrm{A}}}$, $\boldsymbol{p}_j \triangleq [p_j]_{j\in\mathcal{K}_{\mathrm{N}}}$, and $\boldsymbol{v} \triangleq [v_1,\ldots,v_N]^T$ represents the PA locations along the dielectric waveguide. Constraints \eqref{eq:aircomp_power_constraint} and \eqref{eq:noma_power_constraint} impose individual transmit power budgets of the AirComp and NOMA users, respectively. Constraint \eqref{eq:noma_rate_constraint} guarantees the quality-of-service (QoS) requirements of the NOMA users, where $R_{\min,j}$ denotes the minimum achievable rate of user $j$. The AirComp accuracy requirement is captured by \eqref{eq:aircomp_mse_constraint}, which constrains the aggregation MSE to be no larger than a predefined threshold. Finally, constraint \eqref{eq:pa_constraint} ensures that all PAs are deployed within the physical extent of the dielectric waveguide.

Problem \eqref{eq:opt_problem_final} is non-convex due to coupled variables and nonlinear dependence of both the hybrid objective function and the QoS constraints on the PAS-aided equivalent channels. This motivates the development of efficient joint optimization algorithms for precoding design, PA placement, and receive combining, which will be addressed in the subsequent sections.

\section{Proposed Solution}

We adopt an AO framework that decomposes the original problem into tractable subproblems. Specifically, the optimization variables are partitioned into three blocks: transmit powers, receive scalar, and PA location vector. Each block is optimized while fixing the remaining variables.

\subsection{Transmit Power Allocation}

For given $w$ and $\boldsymbol{v}$, define the equivalent PAS-assisted channels as $\bar g_k \triangleq g_k(\boldsymbol{v})$. Due to SIC decoding order, we first optimize NOMA transmit powers and then AirComp powers.

\subsubsection{Power Allocation of NOMA Users}

For fixed $\boldsymbol{p}_k$, the NOMA sum-rate can be rewritten using the telescoping property of SIC as
\begin{equation}
R_{N}
=
B \log_2 \left(
1 +
\frac{\sum_{j \in \mathcal{K}_N} |p_j|^2 |\bar g_j|^2}
{\sum_{k \in \mathcal{K}_A} |p_k|^2 |\bar g_k|^2 + N\sigma^2}
\right).
\end{equation}

Let $\rho_j = |p_j|^2$ and define $I_A = \sum_{k \in \mathcal{K}_A} |p_k|^2 |\bar g_k|^2$. The QoS constraint $R_j^{\mathrm{N}} \ge R_j^{\min}$ can be equivalently rewritten as
\begin{equation}
\rho_j |\bar g_j|^2 
\ge 
\gamma_j 
\left(
\sum_{t=j+1}^{K_N} \rho_t |\bar g_t|^2
+ I_A + N\sigma^2
\right),
\end{equation}
where $\gamma_j = 2^{R_{\min, j}/B} - 1$.
Therefore, the NOMA power allocation subproblem is given by
\begin{subequations}
\begin{align}
\small
\max_{\{\rho_j\}} \quad
& (1-\alpha) B \log_2
\left(
1 +
\frac{\sum_{j} \rho_j |\bar g_j|^2}
{I_A + N\sigma^2}
\right) \\
\text{s.t.}\quad
& \rho_j |\bar g_j|^2 
\ge 
\gamma_j 
\left(
\sum_{t=j+1}^{K_N} \rho_t |\bar g_t|^2
+ I_A + N\sigma^2
\right), \\
& 0 \le \rho_j \le P_j, \quad \forall j \in \mathcal{K}_N .
\end{align}
\end{subequations}
The objective function is concave and all constraints are linear in $\{\rho_j\}$, hence this problem is convex and can be efficiently solved by the standard convex optimization tools such as CVX \cite{grant2014cvx}. The optimal transmit powers are recovered as $p_j^\star = \sqrt{\rho_j^\star}$.

\subsubsection{Power Allocation of AirComp Users}

With $\boldsymbol{p}_j^\star$ obtained, the AirComp power allocation problem is considered. Although the feasible set is convex, the objective function is non-concave. 
To address this issue, we exploit its difference-of-convex (DC) structure and solve it within the DC programming framework \cite{8952884}.
Define
\begin{align}
\small
F(\boldsymbol{p}_k) 
&= (1-\alpha) B \log_2 \!\left(
\sum_{k \in \mathcal{K}_A} |p_k|^2 |\bar g_k|^2 
+ I_N + N\sigma^2
\right) \nonumber \\
& + \alpha B \log_2 \!\left(
\sum_{k \in \mathcal{K}_A} |w^H \bar g_k p_k|^2 
+ N\sigma^2 |w|^2
\right),\\
G(\boldsymbol{p}_k,\beta)
&= (1-\alpha) B \log_2 \!\left(
\sum_{k \in \mathcal{K}_A} |p_k|^2 |\bar g_k|^2 
+ N\sigma^2
\right) \nonumber \\
&\quad + \alpha B \log_2 \!\left(
\beta + N\sigma^2 |w|^2
\right),
\end{align}
where 
$I_N = \sum_{j \in \mathcal{K}_N} |p_j^\star|^2 |\bar g_j|^2$
is constant. 
An auxiliary variable $\beta$ is introduced to satisfy
$\beta \ge 
\sum_{k \in \mathcal{K}_A} |w^H \bar g_k p_k - 1|^2$,
which preserves convexity of the feasible set.
Then, the hybrid objective can be rewritten as
$R_H(\boldsymbol{p}_k) = F(\boldsymbol{p}_k) - G(\boldsymbol{p}_k,\beta)$,
which admits a DC decomposition. 
At iteration $\ell$, the convex function $G(\boldsymbol{p}_k,\beta)$ is approximated by its first-order Taylor expansion at $(\boldsymbol{p}_k^{(\ell)},\beta^{(\ell)})$, resulting in a convex problem that can be efficiently solved. 
By iteratively updating the solution, the DC algorithm generates a sequence of feasible points with non-decreasing objective values and converges to a stationary point.

\subsection{Receive Scalar Optimization}

For fixed transmit powers and PA locations, the receive scalar optimization is given by
\begin{equation}
\max_{w}
\log_2
\left(
\frac{\sum_{k} |w^H \bar g_k p_k|^2 + N\sigma^2 |w|^2}
{\sum_{k} |w^H \bar g_k p_k - 1|^2 + N\sigma^2 |w|^2}
\right)
\end{equation}
subject to the MSE constraint.

Introduce $\tilde w = 1/w$. The problem is equivalent to
\begin{subequations}
\begin{align}
\min_{\tilde w} \quad 
& \sum_{k \in \mathcal{K}_A}
\left|
\tilde w - \bar g_k p_k
\right|^2 \\
\text{s.t.} \quad 
& \sum_{k \in \mathcal{K}_A}
\left|
\tilde w - \bar g_k p_k
\right|^2
+ N\sigma^2
\le
\varepsilon_0 |\tilde w|^2 .
\end{align}
\end{subequations}
The objective is convex in $\tilde w$, whereas the constraint is non-convex due to the quadratic term $|\tilde w|^2$ on the right-hand side. 
To address this issue, we adopt successive convex approximation (SCA) by linearizing $|\tilde w|^2$ via its first-order Taylor expansion at the current iterate. 
The resulting problem is convex and can be efficiently solved at each iteration, yielding a sequence that converges to a stationary point. The optimal receive scalar is recovered as $w^\star = 1/\tilde w^\star$.




\subsection{Optimization of PA Locations}

In the proposed system, the equivalent channels $g_k(\boldsymbol{v})$ vary continuously with the PA locations $\boldsymbol{v}$; however, the non-convex dependence induced by pathloss and phase accumulation renders problem \eqref{eq:opt_problem_final} intractable with respect to $\boldsymbol{v}$, motivating the adoption of a projected gradient ascent (PGA) method \cite{11251272}.

\subsubsection{Projected Gradient Solution} The hybrid rate $R_{{H}}$ depends on $\boldsymbol{v}$ through the equivalent channels
\begin{equation}
g_u(\boldsymbol{v})
=
\sum_{n=1}^{N}
h_{u,n}(v_n)
e^{-j\frac{2\pi n_{\mathrm{ref}}}{\lambda} v_n}.
\end{equation}
Let $r_{u,n} = \|\boldsymbol{u}_k - \boldsymbol{v}_n\|$. The derivative of $g_k$ with respect to $v_n$ is obtained via the chain rule as
\begin{align}
\frac{\partial g_u}{\partial v_n}
&=
\left(
\frac{\partial h_{u,n}}{\partial v_n}
-
j\frac{2\pi n_{\mathrm{ref}}}{\lambda} h_{u,n}
\right)
e^{-j\frac{2\pi n_{\mathrm{ref}}}{\lambda} v_n},
\end{align}
where $\frac{\partial h_{u,n}}{\partial v_n} =h_{u,n}
\left(-\frac{v_n-x_u}{r_{u,n}^2}
-
j\frac{2\pi}{\lambda}
\frac{v_n-x_u}{r_{u,n}}
\right)$.
Using the above expression, the gradient of $R_{{H}}$ with respect to $v_n$ is computed as
\begin{equation}
\frac{\partial R_{{H}}}{\partial v_n}
=
\alpha
\frac{\partial R_{\mathrm{A}}}{\partial v_n}
+
(1-\alpha)
\sum_{j=1}^{K_{\mathrm{N}}}
\frac{\partial R_j^{\mathrm{N}}}{\partial v_n},
\end{equation}
where each term is obtained via standard differentiation of logarithmic rate functions with respect to $|g_k|^2$.
Then, we update the PA locations via PGA method:
\begin{equation}
v_n^{(t+1)}
=
\Pi_{[0,L_x]}
\left(
v_n^{(t)}
+
\eta^{(t)}
\frac{\partial R_H}{\partial v_n}
\right),
\end{equation}
where $\eta^{(t)}$ is determined using backtracking line search and 
$\Pi_{[0,L_x]}(\cdot)$ denotes projection onto $[0,L_x]$.

Since $R_H(\boldsymbol {v})$ is continuously differentiable and the feasible set is compact and convex, the PGA method with appropriate step sizes generates a sequence whose limit points satisfy the first-order KKT conditions. Hence, the algorithm converges to a stationary solution.

\subsection{Convergence and Complexity Analysis}

\subsubsection{Convergence}

Let $R_H^{(t)}$ denote the objective value at the $t$-th iteration. Since each block update does not decrease the objective, the sequence $\{R_H^{(t)}\}$ is monotonically non-decreasing. Moreover, the objective is upper bounded owing to the finite transmit power constraints and the bounded PA deployment region. Therefore, $\{R_H^{(t)}\}$ converges. By standard block coordinate descent arguments, the proposed AO algorithm converges to a stationary point of problem~\eqref{eq:opt_problem_final}.

\subsubsection{Complexity}

The convex NOMA power allocation with $K_N$ variables requires $\mathcal{O}(K_N^3)$ complexity via interior-point methods. 
The AirComp power allocation is solved using DC programming, leading to $\mathcal{O}(T_{\mathrm{DC}} K_A^3)$ complexity, where $T_{\mathrm{DC}}$ is the number of DC iterations. 
The receive scalar update incurs negligible cost, while the PA location optimization requires $\mathcal{O}(T_v K N)$ operations with $T_v$ projected gradient iterations. 
Overall, with $T_{\mathrm{AO}}$ outer iterations, the total complexity is
$
\mathcal{O}
\!\left(
T_{\mathrm{AO}}
\left(
K_N^3
+
T_{\mathrm{DC}} K_A^3
+
T_v K N
\right)
\right)$.

\section{Numerical Results}

In this section, we evaluate the performance of the proposed PAS-assisted hybrid NOMA–AirComp system. 
The system parameters are set as follows \cite{11251272, 11133436}. 
The carrier frequency is $f_c = 28$ GHz, corresponding to wavelength $\lambda = 0.1$ m. 
The system bandwidth is $B = 1$ MHz. 
The noise power is assumed to be $\sigma^2 = -90~\mathrm{dBm}$. 
The dielectric waveguide length is $L_x = 10$ m and deployed at height $d = 5$ m. 
Users are uniformly distributed over a rectangular area of size $10 \times 10$ m$^2$. 
Unless otherwise stated, we set $K_A = 4$, $K_N = 3$, $N = 6$, $\alpha = 0.5$, and identical maximum transmit powers $P_k = P_j = P_{\text{max}} =20$ dBm. 
The minimum rate requirement of each NOMA user is $R_{\min, j}=0.5$ Mbps and the AirComp MSE threshold is $\varepsilon_0 = 10^{-2}$. All results are averaged over 200 independent user location realizations.
We compare the proposed design with the following benchmark schemes:

\begin{itemize}
    \item \textbf{Fixed PA position:} 
    PAs are uniformly distributed on the waveguide with fixed locations given by $v_{n} = \frac{nL_x}{N+1}$. Only transmit powers and receive scalar are optimized.
    \item \textbf{Discrete PAS:} 
    Each antenna can only be selected from a predefined discrete grid of positions along the waveguide.

    \item \textbf{Full power transmission:} 
    All users transmit at their maximum power $P_{\text{max}}$, while PA locations and receive scalar are optimized. 

\end{itemize}

\begin{figure*}[!t]
\centering
\subfloat[\scriptsize]{%
\includegraphics[width=0.25\textwidth,height=3.3cm]{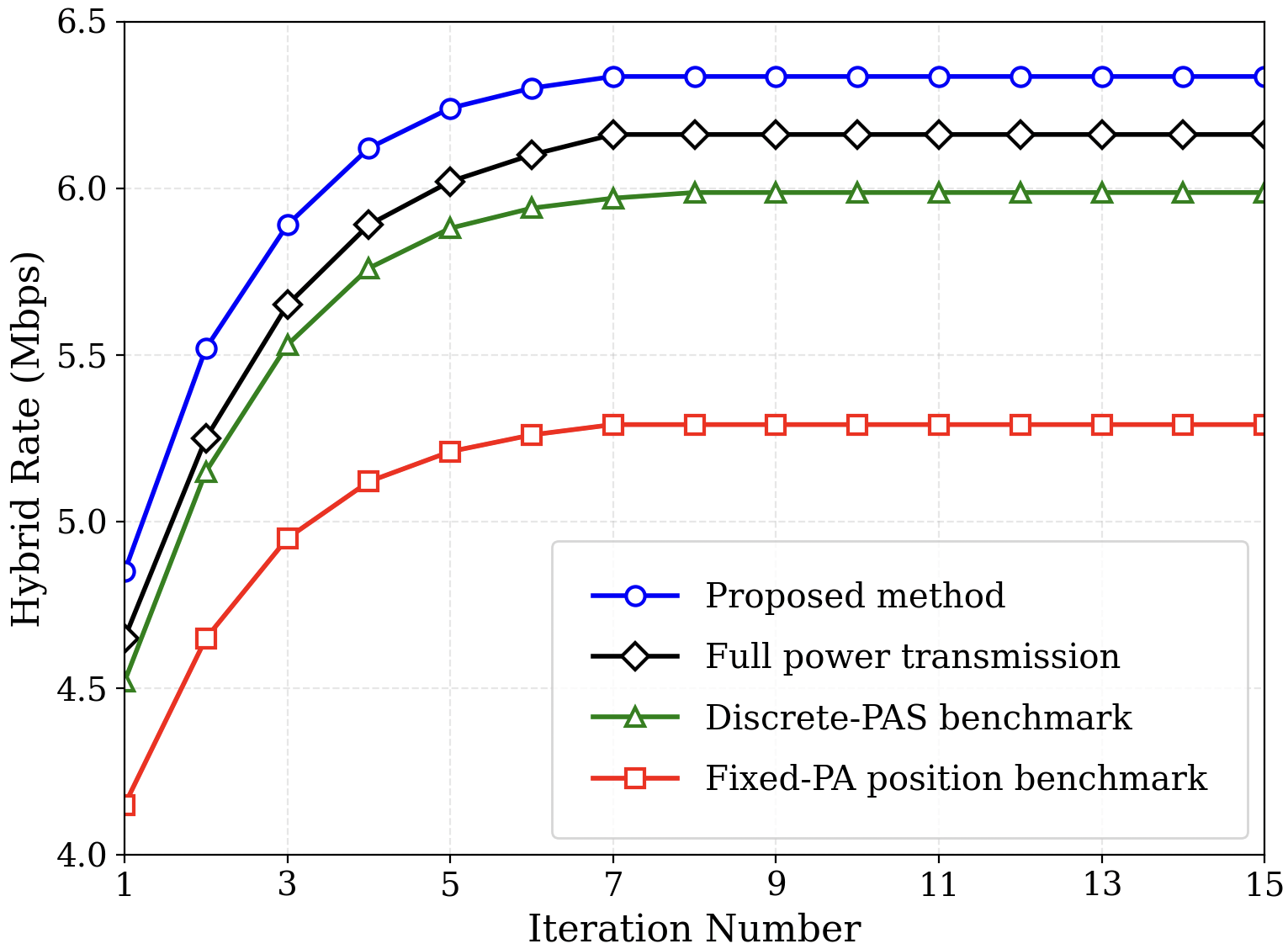}%
    \label{fig:iteration}}
\hfil
\subfloat[\scriptsize]{%
\includegraphics[width=0.25\textwidth,height=3.3cm]{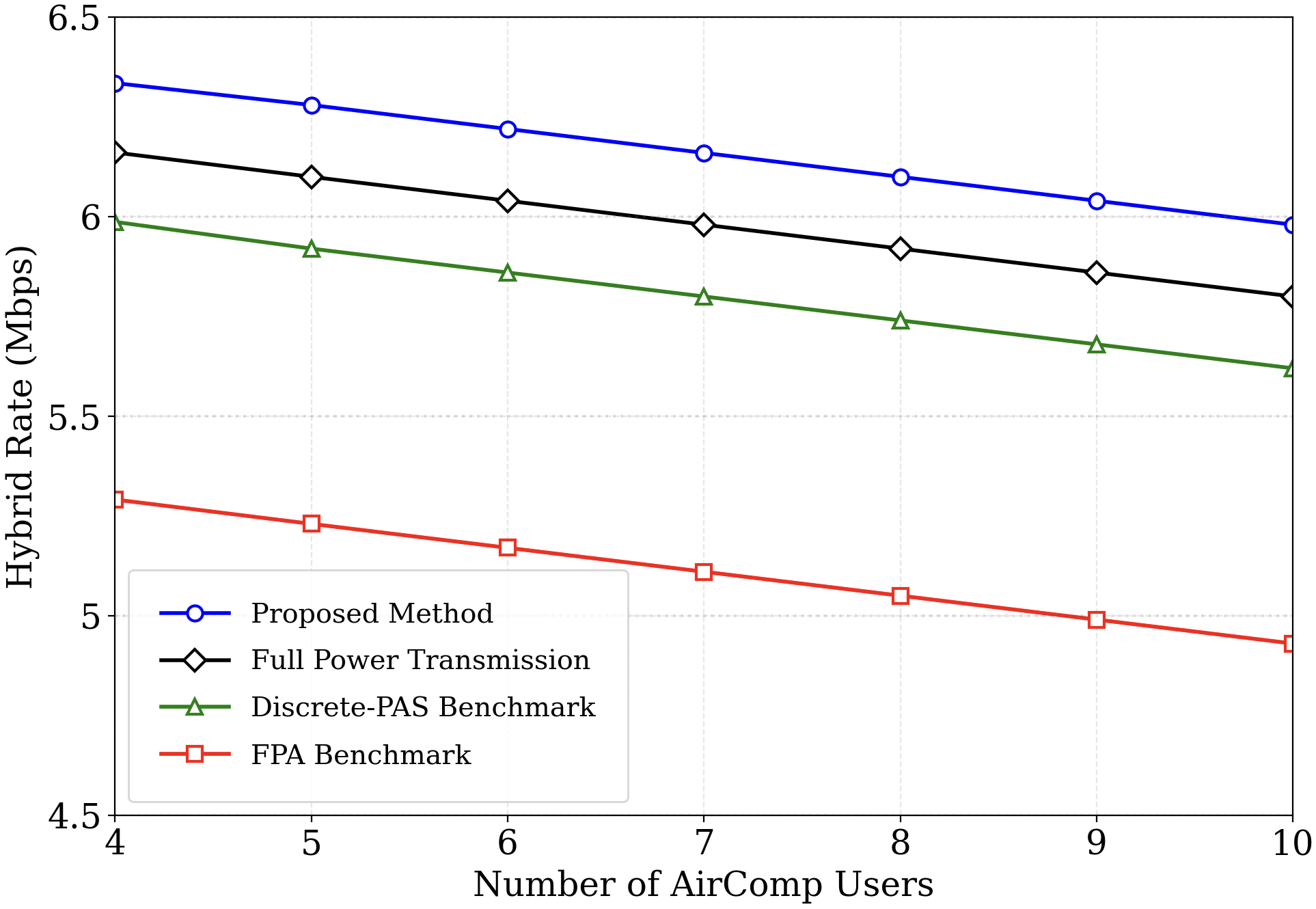}%
\label{fig:aircomp_users}}
\hfil
\subfloat[\scriptsize]{%
\includegraphics[width=0.25\textwidth,height=3.3cm]{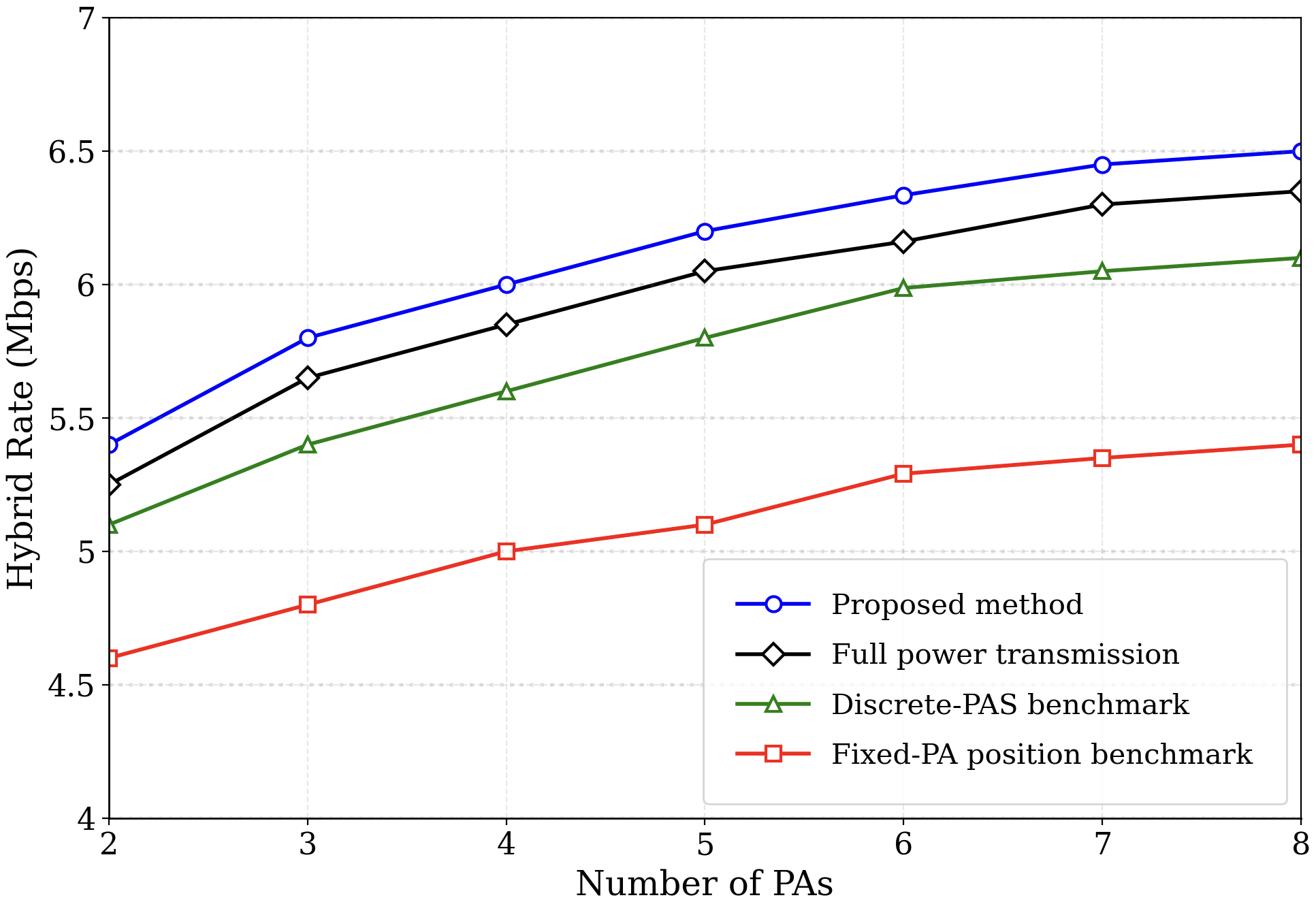}%
    \label{fig:number_pa}}
\hfil
\subfloat[\scriptsize]{%
\includegraphics[width=0.25\textwidth,height=3.3cm]{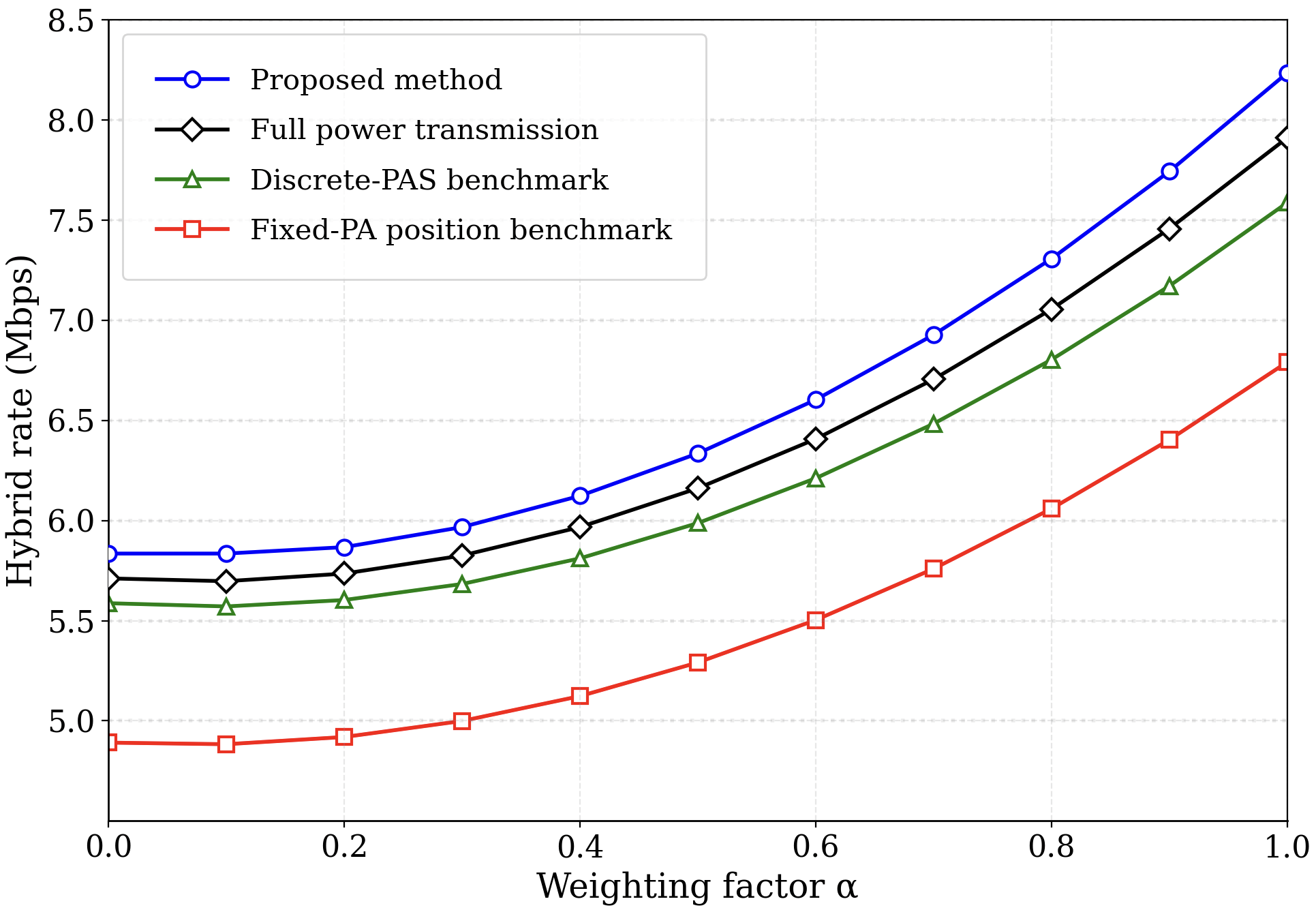}
    \label{fig:alpha_variation}}
\caption{(a) Convergence performance of our proposed design. (b) Hybrid rate w.r.t. the number of AirComp users. (c) Hybrid rate w.r.t. the number of PAs. (d) Hybrid rate versus the weighting factor $\alpha$.}
\label{fig:combined}
\end{figure*}

Fig.~\ref{fig:iteration} illustrates the convergence behavior of hybrid rate versus the number of AO iterations. It can be observed that the hybrid rate increases rapidly with the number of iterations and converges within approximately 8 iterations. Moreover, the proposed method consistently achieves a higher hybrid rate compared with the benchmark schemes, demonstrating the effectiveness of the proposed joint optimization framework.

Fig.~\ref{fig:aircomp_users} shows the hybrid rate as a function of the number of AirComp users $K_A$, while fixing $K_N=3$ and $N=6$. 
As $K_A$ increases, more computation signals share the channel, leading to stronger AirComp interference at the NOMA decoding stage and increased aggregation distortion. Therefore, the hybrid rate gradually decreases for all schemes. Nevertheless, the proposed PAS design consistently achieves a higher hybrid rate than the benchmark schemes, due to the additional spatial DoF that enable more effective shaping of the equivalent channels.


Fig.~\ref{fig:number_pa} presents the hybrid rate versus the number of pinching antennas $N$, with $K_A=4$ and $K_N=3$. It is observed that increasing $N$ improves the hybrid rate for all schemes due to enhanced spatial diversity and array gain. However, the performance gap between the proposed design and the benchmark schemes enlarges as $N$ increases. This is because a larger number of antennas provides more spatial DoF, which the proposed continuous optimization fully exploits, whereas the fixed-position scheme cannot adapt its geometry and D-PAS is constrained by discrete positioning resolution.

Fig.~\ref{fig:alpha_variation} illustrates the impact of the weighting factor $\alpha$ on the hybrid rate, which governs the tradeoff between AirComp computation rate and NOMA sum-rate. As $\alpha$ increases, the system progressively shifts from a communication-oriented regime toward a computation-oriented regime. In the low-$\alpha$ region, the hybrid performance is dominated by the NOMA component, resulting in relatively similar performance across all schemes. However, as $\alpha$ increases, the proposed PAS-assisted design achieves noticeably higher hybrid rates compared to the benchmarks. This gain stems from its ability to jointly optimize antenna placement and precoding, thereby effectively shaping the equivalent channels to reduce aggregation distortion while controlling interference.


\section{Conclusion}

This paper investigated a PAS-assisted hybrid framework integrating AirComp and NOMA. A hybrid rate maximization problem was formulated by jointly optimizing transmit power, the AirComp receive scalar, and the positions of pinching antennas. An AO algorithm was developed to solve the resulting non-convex problem. Simulation results demonstrate that the proposed design achieves significant performance improvements over benchmark schemes.

\bibliographystyle{IEEEtran}
\bibliography{refpinching}

\vfill
	
\end{document}